\documentclass[a4paper,UKenglish,cleveref, autoref, thm-restate]{lipics-v2021}

\hideLIPIcs  

\title{\textsf{weberknecht} -- a \OSCM{} solver}

\author{Johannes Rauch}
{Ulm University, Institute of Optimization and Operations Research, Germany}
{johannes.rauch@uni-ulm.de}
{https://orcid.org/0000-0002-6925-8830}
{}

\authorrunning{J. Rauch} 

\Copyright{Johannes Rauch} 



\supplement{\href{https://doi.org/10.5281/zenodo.12154892}{\texttt{https://doi.org/10.5281/zenodo.12154892}},\\ \href{https://github.com/johannesrauch/PACE-2024}{\texttt{https://github.com/johannesrauch/PACE-2024}},
\href{https://pacechallenge.org/2024/}{\texttt{https://pacechallenge.org/2024/}}
}

\acknowledgements{I thank Henning Bruhn-Fujimoto and Dieter Rautenbach for helpful discussions.}

\nolinenumbers 

\EventEditors{Philipp Kindermann and Fabian Klute and Soeren Terziadis}
\EventNoEds{3}
\EventLongTitle{Parameterized Algorithms and Computational Experiments 2024}
\EventShortTitle{PACE 2024}
\EventAcronym{PACE}
\EventYear{2024}
\EventDate{2023--2024}
\EventLocation{}
\EventLogo{}
\SeriesVolume{$\times$}
\ArticleNo{$\times$}

\usepackage{tikz}
\tikzset{
	dot/.style = {circle, fill, minimum size=#1,
		inner sep=0pt, outer sep=0pt},
	dot/.default = 5pt
}

\usepackage{lmodern}

\newcommand{\OSCM}{\textsc{One-Sided Crossing Minimization}}

\begin{document}

\maketitle

\begin{abstract}
We describe the implementation of the exact solver \textsf{weberknecht}\footnote{Weberknecht is the german name for the harvestman spider. It is a composite word consisting of the words Weber = weaver and Knecht = workman. 
} and the heuristic solver \textsf{weberknecht\_h} for the \OSCM{} problem.
\end{abstract}

\section{Introduction and preliminaries}\label{sec:pre}
An instance $(G = (A \dot{\cup} B, E), \pi_A)$ of \OSCM{} is a bipartite graph $G$ with $n$ vertices, bipartition sets $A$ and $B$, and a linear ordering $\pi_A$.
The goal is to find a linear ordering $\pi_B$ of $B$ that minimizes the number of crossing edges if the graph were to be drawn in the plane such that
the vertices of $A$ and $B$ are on two distinct parallel lines, respectively,
the order of the vertices of $A$ and $B$ on the lines is consistent with the linear orderings $\pi_A$ and $\pi_B$, respectively, and
the edges are drawn as straight lines.
\OSCM{} was the Parameterized Algorithms and Computational Experiments Challenge 2024\footnote{\href{https://pacechallenge.org/2024/}{\texttt{https://pacechallenge.org/2024/}}}.

We may assume that $A = [n_0] := \{1, \dots, n_0\}$ and $B = \{n_0 + 1, \dots, n_0 + n_1\}$ for some positive integers $n_0$ and $n_1$.
We think of $\pi_A$ and $\pi_B$ as bijections $A \rightarrow [n_0]$ and $B \rightarrow [n_1]$, respectively.
With this we can view \OSCM{} as a purely combinatorial problem, that is, the edges $a_1b_1$ and $a_2b_2$ of $G$ with $a_1,a_2 \in A$ and $b_1,b_2 \in B$ cross if and only if
\begin{center}
($\pi_A(a_1) < \pi_A(a_2)$ and $\pi_B(b_1) > \pi_B(b_2)$) or ($\pi_A(a_1) > \pi_A(a_2)$ and $\pi_B(b_1) < \pi_B(b_2)$),
\end{center}
and we wish to minimize the number of crossing edges.
If $\pi_B(u) < \pi_B(v)$ for $u, v \in B$, we say that $u$ is ordered before $v$, or $u$ is to the left of $v$.

Let $c_{u,v}$ denote the number of crossings of edges incident to $u, v \in B$ with $u \neq v$ if $\pi_B(u) < \pi_B(v)$.
A mixed-integer program for \OSCM{} is given by
\begin{equation}
\label{mip}
\begin{alignedat}{2}
\text{minimize } &\sum_{\substack{u, v \in B\\u < v}} (c_{u,v} - c_{v,u}) \cdot x_{u,v} + \sum_{\substack{u, v \in B\\u < v}} c_{v,u}\\
\text{subject to } &\begin{aligned}[t]
0 \leq x_{u,v} + x_{v,w} - x_{u,w} \leq 1 &\quad\text{for all }u,v,w \in B, u < v < w,\\
x_{u,v} \in \{0,1\} &\quad\text{for all }u,v \in B, u < v.
\end{aligned}
\end{alignedat}\tag{$P_I$}
\end{equation}
We refer to the article of Jünger and Mutzel~\cite{juenger1997twolayer} for a detailed derivation of the mixed-integer program (\ref{mip}).
Observe that $u$ is ordered before $v$ if and only if $x_{u,v} = 1$ for $u, v \in B$ with $u < v$.

For more information on \OSCM{} and graph drawing in general we refer to the handbook of graph drawing and visualization of Tamassia~\cite{tamassia2013handbook}.

\section{Overview}
We describe the general modus operandi of \textsf{weberknecht(\_h)}, which are written in C++ and available on GitHub\footnote{\href{https://github.com/johannesrauch/PACE-2024/}{\texttt{https://github.com/johannesrauch/PACE-2024/}}}.
First, the exact solver \textsf{weberknecht} runs the uninformed and improvement heuristics described in Section~\ref{sec:heu}.
Then it applies the data reduction rules described in Section~\ref{sec:red}.
Last, it solves a reduced version of the mixed-integer program (\ref{mip}) associated to the input instance with a custom branch and bound and cut algorithm described in Section~\ref{sec:bnb}.
The heuristic solver \textsf{weberknecht\_h} only runs the uninformed and improvement heuristics (except the local search heuristic).

\section{Heuristics}\label{sec:heu}
We distinguish between uninformed and informed heuristics, which build a solution from the ground up, and improvement heuristics, which try to improve a given solution.
Due to the reduction rules we may assume from here that there are no isolated vertices in $G$.

\medskip
\noindent
\textbf{Uninformed Heuristics.}
The uninformed heuristics order the vertices of $B$ such that the \emph{scores} $s(v)$ of vertices $v \in B$ is non-decreasing.
\begin{itemize}
\item In the \emph{barycenter heuristic}, we set $s(v) = \frac{1}{d_G(v)} = \sum_{u \in N_G(v)} u$ (recall that $A = [n_0]$). Eades and Wormald~\cite{eades1994edge} proved that this method has an $\mathcal{O}(\sqrt{n})$ approximation factor, which is best possible up to a constant factor under certain assumptions.
\item Let $d = d_G(v)$ and let $w_0, \dots, w_{d-1}$ be the neighbors of $v$ in $G$ with $w_0 < \dots < w_{d-1}$.
In the \emph{median heuristic}, the score of $v$ is $s(v) = w_{(d-1)/2}$ if $d$ is odd and $s(v) = (w_{d/2 - 1} + w_{d/2})/2$ if $d$ is even.
Eades and Wormald~\cite{eades1994edge} proved that this method is a 3-approximation algorithm.
\item In the \emph{probabilistic median heuristic}, we draw a value $x$ from $[0.0957, 0.9043]$ uniformly at random, and the score of $v$ is then $s(v) = w_{\lfloor x \cdot d \rfloor}$.
This is essentially the approximation algorithm of Nagamochi~\cite{nagamochi2005improved}, which has an approximation factor of $1.4664$ in expectancy.
\end{itemize}

\noindent
\textbf{Informed Heuristics.}
The informed heuristics get a fractional solution of the linear program relaxation of \eqref{mip} as an additional input.
\begin{itemize}
\item The \emph{sort heuristic} works like a uninformed heuristics.
The score for vertex $v \in B$ is $s(v) = \sum_{u \in B, u < v} x_{u,v} + \sum_{u \in B, v < u} (1 - x_{v, u})$.
\item Classical \emph{randomized rounding heuristic}.
\item \emph{Relaxation induced neighborhood search}~\cite{danna2005exploring}.
\end{itemize}

\noindent
\textbf{Improvement Heuristics.}
Assume that $\pi_B = u_1u_2 \dots u_{n_1}$ is the current best solution.
\begin{itemize}
\item The \emph{shift heuristic} that Grötschel et al.~\cite{grotschel1984cutting} describe tries if shifting a single vertex improves the current solution.
\item In the \emph{local search heuristic}, we try to solve a reduced version of \eqref{mip} to optimality, where we only add variables $x_{u_i, u_j}$ with $|i-j| < w$ for some parameter $w$.
\end{itemize}

\section{Data reduction}\label{sec:red}
The solver \textsf{weberknecht} implements the following data reduction rules.
\begin{itemize}
\item Vertices of degree zero in $B$ are put on the leftmost positions in the linear ordering $\pi_B$.
Note that there is an optimal solution with exactly these positions for the isolated vertices in $B$.
\item Let $l_v$ ($r_v$) be the neighbor of $v \in B$ in $G$ that minimizes (maximizes) $\pi_A$, respectively.
Dujmovi\'c and Whitesides~\cite{dujmovic2004efficient} noted that, if there exists two nonempty sets $B_1, B_2 \subseteq B$ and a vertex $q \in A$ such that for all $v \in B_1$ we have that $\pi_A(r_v) \leq \pi_A(q)$, and for all $v \in B_2$ we have that $\pi_A(q) \leq \pi_A(l_v)$,
then the vertices of $B_1$ appear before the vertices of $B_2$ in an optimal solution.
In this case we can split the instance into two subinstances.
\item Dujmovi\'c and Whitesides~\cite{dujmovic2004efficient} proved that, 
if $\pi_B$ is an optimal solution, and $c_{u, v} = 0$ and $c_{v, u} > 0$,
then $\pi_B(u) < \pi_B(v)$.
\item Dujmovi\'{c} et al.~\cite{dujmovic2008fixed} described a particular case of the next reduction rule.
Let $c_{u, v} < c_{v, u}$. 
We describe the idea with the example in Figure~\ref{fig:rr}.
Imagine that we draw some edge $x_iy_j$ into Figure~\ref{fig:rr}.
If the number of edges crossed by $x_iy_j$ on the left side is at most the number of edges crossed by $x_iy_j$ on the right side for all edges of the form $x_iy_j$, then we have $\pi_B(u) < \pi_B(v)$ in any optimal solution $\pi_B$:
Otherwise we could improve the solution by simply exchanging the positions of $u$ and $v$.
Note that this reduction rule is only applicable if $d_G(u) = d_G(v)$ as witnessed by $x_2y_1$ and $x_2y_k$ ($k=5$ here).
\item The value $\ell b = \sum_{u, v \in B, u < v} \min(c_{u,v}, c_{v,u})$ is a lower bound on the number of crossings of an optimal solution.
Suppose that we have already computed a solution with $ub$ crossings.
Then, if $c_{u,v} \geq ub - \ell b$ for some $u, v \in B$, it suffices to only consider orderings $\pi_B$ with $\pi_B(u) > \pi_B(v)$ for the remaining execution.
\item After the execution of the described reduction rules, some variables $x_{u,v}$ of (\ref{mip}) have a fixed value due to the constraints of (\ref{mip}).
\end{itemize}

\begin{figure}
\centering
\begin{tikzpicture}
\begin{scope}[xscale=0.5, shift={(-7,0)}]
\node[dot=2, label=below:$x_1$] (x1) at (-2,0) {};
\node[dot, label=below:$u$] (u) at (-1,0) {};
\node[dot=2, label=below:$x_2$] (x2) at (0,0) {};
\node[dot, label=below:$v$] (v) at (1,0) {};
\node[dot=2, label=below:$x_3$] (x3) at (2,0){};

\node[dot=2, label=above:$y_1$] at (-4,1) {};
\node[dot] (n1) at (-3,1) {};
\node[dot=2, label=above:$y_2$] at (-2,1) {};
\node[dot] (n2) at (-1,1) {};
\node[dot=2, label=above:$y_3$] at (0,1) {};
\node[dot] (n3) at (1,1) {};
\node[dot=2, label=above:$y_4$] at (2,1) {};
\node[dot] (n4) at (3,1) {};
\node[dot=2, label=above:$y_5$] at (4,1) {};

\draw (n1) to (u);
\draw (n3) to (u);
\draw (n4) to (u);
\draw (n2) to (v);
\draw (n3) to (v);
\draw (n4) to (v);
\end{scope}
\begin{scope}[xscale=0.5, shift={(7,0)}]
\node[dot=2, label=below:$x_1$] (x1) at (-2,0) {};
\node[dot, label=below:$u$] (u) at (1,0) {};
\node[dot=2, label=below:$x_2$] (x2) at (0,0) {};
\node[dot, label=below:$v$] (v) at (-1,0) {};
\node[dot=2, label=below:$x_3$] (x3) at (2,0){};

\node[dot=2, label=above:$y_1$] at (-4,1) {};
\node[dot] (n1) at (-3,1) {};
\node[dot=2, label=above:$y_2$] at (-2,1) {};
\node[dot] (n2) at (-1,1) {};
\node[dot=2, label=above:$y_3$] at (0,1) {};
\node[dot] (n3) at (1,1) {};
\node[dot=2, label=above:$y_4$] at (2,1) {};
\node[dot] (n4) at (3,1) {};
\node[dot=2, label=above:$y_5$] at (4,1) {};

\draw (n1) to (u);
\draw (n3) to (u);
\draw (n4) to (u);
\draw (n2) to (v);
\draw (n3) to (v);
\draw (n4) to (v);
\end{scope}
\end{tikzpicture}
\caption{}
\label{fig:rr}
\end{figure}

\section{Branch and bound}\label{sec:bnb}
The solver \textsf{weberknecht} implements a rudimentary branch and bound algorithm.
We use HiGHS~\cite{huangfu2018parallelizing} only as a linear program solver, and not as a mixed-integer program solver, since the mixed-integer program solver does not (yet) implement lazy constraints.
To avoid adding all $\Theta(n^3)$ constraints, we solve the linear program relaxation of \eqref{mip} as follows.
\begin{enumerate}[1.]
\item Create a linear program $(P)$ with the objective function of \eqref{mip} and no constraints.
\item Solve $(P)$.
\item If the current solution violates constraints of \eqref{mip}, add them to $(P)$ and go to 2.
\end{enumerate}
Let $ub$ denote the number of crossings of the current best solution.
Then, until we have a optimal solution, \textsf{weberknecht} executes the following.
\begin{enumerate}[1.]
\item Solve $(P)$ with the method described above.
\item If $(P)$ is infeasible, backtrack.
\item If the rounded objective value of $P$ is at least $ub$, backtrack.
\item If the current solution of $(P)$ is integral, update the best solution and backtrack.
\item Run informed heuristics and branch.
\end{enumerate}

\end{document}